\documentclass[preprint,11pt]{elsarticle}
\usepackage[latin1]{inputenc}
\usepackage[T1]{fontenc}
\usepackage [english]{babel}
\usepackage{graphicx}
\usepackage{subfigure}
\usepackage{epsfig}
\usepackage{amssymb}
\usepackage{amsmath}
\usepackage{color}
\usepackage{fancyhdr}
\usepackage{amstext}
\usepackage{amsbsy}
\usepackage{amsfonts}
\usepackage{lineno}

\journal{The Astrophysical Journal}

\begin{document}

\begin{frontmatter}

\title{The \textit{East-West method}: an exposure-independent method \\
to search for large scale anisotropies of cosmic rays}

\renewcommand{\thefootnote}{\fnsymbol{footnote}}

\author[1,2]{R.Bonino}
\author[3]{V.V.Alekseenko} 
\author[4]{O.Deligny}
\author[1,5]{P.L.Ghia}
\author[6]{M.Grigat}
\author[5] {A.Letessier-Selvon}
\author[4,7,8]{H.Lyberis} 
\author[9]{S.Mollerach}
\author[10]{S.Over}
\author[9]{E.Roulet}

\address[1]{Istituto di Fisica dello Spazio Interplanetario-INAF, Torino, Italy}
\address[2]{Istituto Nazionale di Fisica Nucleare, Torino, Italy}
\address[3]{Institute for Nuclear Research, AS Russia, Baksan Neutrino Observatory, Russia}
\address[4]{IPNO, Universit\'e Paris Sud \& CNRS-IN2P3, Orsay, France}
\address[5]{LPNHE, Universit\'es Paris 6 et Paris 7 \& CNRS-IN2P3, Paris, France}
\address[6]{RWTH Aachen University, III Physikalisches Institut A, Aachen, Germany}
\address[7]{Universit\`a degli Studi di Torino, Torino, Italy}
\address[8]{Universit\'e Paris VII Denis Diderot, Paris, France}
\address[9]{CONICET, Centro At\'omico Bariloche, Bariloche, Argentina}
\address[10]{Universitat Siegen, Siegen, Germany}

\begin{abstract}
The measurement of large scale anisotropies in cosmic ray 
arrival directions at energies above $10^{13}$ eV is performed through 
the detection of Extensive Air Showers produced by cosmic ray
interactions in the atmosphere. The observed anisotropies are small, 
so accurate measurements require small statistical uncertainties,
\textit{i.e.} large datasets. These can be obtained by employing ground
detector arrays with large extensions (from $10^4$ to $10^9$ m$^2$) and
long operation time (up to 20 years). The control of such arrays is 
challenging and spurious variations in the counting rate due to instrumental effects (\textit{e.g.} 
data taking interruptions or changes in the acceptance) and atmospheric effects (\textit{e.g.} air 
temperature and pressure effects on EAS development) are usually 
present. These modulations must be corrected 
very precisely before performing standard anisotropy analyses, \textit{i.e.} 
harmonic analysis of the counting rate versus local sidereal time.   
\noindent In this paper we discuss an alternative method to measure 
large scale anisotropies, the ``East-West method'', originally proposed 
by Nagashima in 1989. 
It is a differential method, as it is based on the 
analysis of the difference of the counting rates in the East and West 
directions. Besides explaining the principle, we present here its 
mathematical derivation, showing that the method is largely independent 
of experimental effects, that is, it does not require corrections for 
acceptance and/or for atmospheric effects. We explain the use of the 
method to derive the amplitude and phase of the anisotropy and we 
demonstrate its power under different conditions of detector operation. 
\end{abstract}

\begin{keyword} Large scale anisotropy \sep First harmonic analysis \sep East-West method
\end{keyword}

\end{frontmatter}

\section{Introduction}

The cosmic ray (CR) spectrum, in spite of its apparent regularity, exhibits in fact
a few features, namely a first change of slope - the ``knee'' - near 
$10^{15}$ eV, a second bending of the spectrum - the ``second knee'' - 
near $10^{17}$ eV, a hardening - ``ankle'' - around $10^{18}$ eV, and 
a flux suppression around $5 \times 10^{19}$ eV. 
The measurement of the anisotropy in the arrival directions of cosmic
rays is a complementary tool, with respect to the energy spectrum 
and mass composition, to investigate the origins of these features. From 
the observational point of view indeed, the study of the CR anisotropy, 
and especially its evolution over the energy spectrum, is closely 
connected to the problem of CR propagation and sources. 

Current experimental 
results show that the main features of the anisotropy are similar in the energy range ($10^{11} \div 10^{14}$~eV), both 
with respect to amplitude ($10^{-4} \div 10^{-3}$) and phase ((0 $\div $ 4)~h
LST)~\cite{Nagashima, musala, poatina, baksan1, baksan2, MACRO, kamioka, tibet, 
superk, milagro,eastop1}. At higher energies, between $10^{14}$ and $10^{17}$~eV, the limited statistics do not 
allow any firm conclusion to be drawn~\cite{akeno, gherardy, kascadean, tibetsci, 
kgrande}, although the observation of a larger anisotropy amplitude with a 
different phase has been recently reported at $\sim4\cdot 10^{14}$~eV~\cite{eastop2}.
Around $10^{18}$~eV, when the gyro-radius of the particles becomes comparable 
to the galactic disk thickness, we would expect 
a large increase in the anisotropy (\emph{i.e.} amplitudes 
at the percent level), in particular towards the galactic disk.  
From the experimental point of view, since the statistics are obviously more limited in this energy range, 
the situation is not as clear as at lower energies~\cite{AGASA,flyseye,sugar,Auger_ICRC,Auger_APP}. 
Above $10^{18}$~eV, CRs (believed to be extra-galactic due to loss of 
confinement in the Galaxy) are not expected to show significant large scale 
modulations, with typical expectations for the amplitude arising from the 
Compton-Getting effect being below the one percent level. The change from a large to an almost 
null amplitude would mark the transition from galactic to extra-galactic CR origin. In few years 
the present experiments (e.g. the Pierre Auger Observatory~\cite{augerdet}) will have collected
enough statistics to provide additional informations to probe this
energy range.

The experimental study of large scale CR anisotropies is thus fundamental for cosmic 
ray physics, though it is challenging. At energies $\geq 10^{13}$~eV such measurements 
have to be indirect (due to too low primary fluxes) and are usually performed 
through Extensive Air Shower (EAS) arrays. These arrays operate almost uniformly with respect to sidereal time
thanks to the Earth's rotation: the zenith angle dependent shower detection and 
reconstruction is not a function of right ascension but it is a strong function of 
declination. Thus, the most commonly used technique (originally proposed in~\cite{Linsley}) is 
the analysis in right ascension only, through harmonic analysis (Rayleigh formalism) 
of the counting rate within the declination band defined by the detector field of view. 
Conventionally, one extracts the first and second harmonic: this is done by measuring 
the counting rate as a function of the local sidereal time (or right ascension), and fitting 
the result to a sine wave. The Rayleigh formalism gives the amplitude of the different 
harmonics, the corresponding phase (right ascension of the maximum intensity) and the 
probability of detecting a signal due to fluctuations of an isotropic distribution with an 
amplitude equal or larger than the observed one. 

The technique in itself is rather simple but the greatest difficulties are in the 
treatment of the data, \textit{i.e.} of the counting rates themselves. Both for 
large scale anisotropies linked to diffusive motions or for the ones due to the 
Compton-Getting effect (\textit{i.e.} due to the observer's motion with respect 
to a locally isotropic population of cosmic rays), the expected amplitudes are very 
small ($10^{-6}-10^{-2}$), with related statistical problems: long term observations 
and large collecting areas are needed. Instrumental effects must be kept as small as 
possible, requiring detectors to operate uniformly (both in size and over time), and 
being as stable as possible. Moreover, EAS arrays are mostly located in remote sites 
(generally at mountain altitude), being thus subject to large atmospheric variations, 
both in temperature and in pressure. Meteorological induced modulations can affect the 
CR rate: indeed the EAS properties themselves depend on air density (through variations 
of the Moli\`ere radius) and on pressure (due to the absorption of the electromagnetic 
component in the air) \cite{Auger_weather}.

The measurement in practice is in fact complicated by the need of correcting the counting 
rate for instrumental and atmospheric effects, that must be done with high precision to 
prevent the introduction of artificial variations in the CR flux. 

The \emph{East-West method}, being based on a differential technique, was  designed to 
avoid introducing such corrections, preventing  the possible associated systematics to 
affect the results. The original idea was proposed in~\cite{Nagashima} and applied 
to the data of the Mt. Norikura array. It was later applied by other EAS arrays such as the 
Tibet experiment~\cite{tibet}, EAS-TOP~\cite{eastop1} and the Pierre Auger 
Observatory~\cite{Auger_ICRC,Auger_APP}. A modification of this technique was employed by the 
Milagro experiment~\cite{milagro}. We revisit here this method, with the idea of explaining 
its mathematical background and of showing its power when applied to EAS data under different 
conditions of operation. In section~\ref{section:method} we explain the principle and 
show that the classical implementation of the method is valid within certain approximations. 
In section~\ref{section:ew} we derive the mathematical basis of the method, avoiding some
approximations used in the classical implementation and meanwhile demonstrating that the 
East-West technique is largely independent of any instrumental/atmospheric effect. 
The derivation of the amplitude and phase of the anisotropy from the harmonic analysis of 
the differences in East and West directions is illustrated in section~\ref{section:fha}. 
Here we show also how to extract from the derived amplitude and phase those corresponding 
to the equatorial component of the dipole. Before concluding in 
section~\ref{section:conclusion}, we apply in section~\ref{section:mc} the East-West
method to different mock EAS data sets, characterized by different spurious 
effects.  

\section{The principle of the East-West method and its classical implementation} \label{section:method}

The East-West method relies on the fact that the difference 
between the observed counting rates of events recorded at each local sidereal time $t$ 
(ranging from 0 to 24 hs ; \textit{i.e.} superimposing all detected events in a unique 
sidereal day) arriving from the Eastern and Western hemispheres, $I_E^{obs}(t)$  
and $I_W^{obs}(t)$ respectively, is proportional to the derivative of the true total 
counting rate $I_{tot}^{true}(t)$, the coefficient of proportionality being approximately 
the mean hour angle $\left<h\right>$ of the observed events. In this section, we aim at retrieving this classical 
implementation by outlining the different approximations which are needed to obtain this 
result. We will derive in the next section the relationship between  
$I_E^{obs}(t)-I_W^{obs}(t)$ and $I_{tot}^{true}(t)$ in a more rigorous way. 

\subsection{The principle}

The total counting rate of events observed in either the Eastern or the Western half of 
the field of view of an EAS array experiences different kind of variations during a sidereal 
day. Those may be caused either by experimental effects (changes of measurement 
conditions during the data taking, atmospheric effects on EAS, etc.) and/or by real 
variations in the primary CR fluxes from different parts of the sky. The East-West method 
is aimed at reconstructing the equatorial component of a genuine large scale pattern by 
using only the difference of the counting rates of the Eastern and Western hemispheres. 
The effects of experimental origin, being independent of the incoming direction, 
are expected to be removed through the subtraction\footnote{This is a natural 
expectation provided the fact that the azimuthal detection efficiency of the corresponding 
experiment is symmetrical in East-West. Note however that corrections due to eventual 
asymmetries (such as those expected if the array is tilted) can be applied, as long as the 
asymmetries are known.}. In the presence of a genuine dipolar distribution 
of CRs, as the Earth rotates Eastwards, the Eastern sky is closer to the dipole excess 
region for half a day each day; then, after the field of view has traversed the excess 
region, the Western sky becomes closer to the excess region and thus bears higher counting 
rates than the Eastern sky. The East-West differential counting rate is thus subject to 
oscillations whose amplitude and phase are expected to be related to those of the genuine 
large scale anisotropy. 

\subsection{Classical implementation of the East-West method}\label{subsec:class_implementation}

We denote the CR flux by $\Phi(\alpha,\delta)=\Phi_0+\Phi_1(\alpha,\delta)$, expressed
in equatorial coordinates and where $\Phi_0$ stands for the isotropic component, which is
by construction large compared to the anisotropic one $\Phi_1$ (defined 
such that its average over the sky vanishes). The true 
counting rates (\textit{i.e.} those that would be measured by a perfectly stable operating experiment and 
with negligible atmospheric effects on shower developments) from the Eastern and Western sectors, 
$I_E^{true}(t)$ and $I_W^{true}(t)$, can be 
expressed in terms of the CR flux $\Phi(\alpha,\delta)$ as: 
\begin{eqnarray}
\label{eqn:rates_true}
I_E^{true}(t)&=&A\int_{\delta_{min}}^{\delta_{max}} \hspace{-0.6cm}d\delta\,\cos\delta\,\int_t^{t+\pi} \hspace{-0.4cm}d\alpha\,\omega(t-\alpha,\delta)\,\Phi(\alpha,\delta),\\
I_W^{true}(t)&=&A\int_{\delta_{min}}^{\delta_{max}} \hspace{-0.6cm}d\delta\,\cos\delta\,\int^t_{t-\pi} \hspace{-0.2cm}d\alpha\,\omega(t-\alpha,\delta)\,\Phi(\alpha,\delta),
\end{eqnarray}
where $A$ is the effective area of the experiment and $t$ stands hereafter, if not otherwise 
specified, for the local sidereal time and varies between 0 and $2\pi$. 
Ignoring spurious modulation effects that will be
considered below, the experimental exposure $\omega(t,\alpha,\delta)$
of a ground experiment as function of the local sidereal time $t$, the right ascension $\alpha$ 
and the declination $\delta$ depends only on the combination $(t-\alpha,\delta)$ in an even 
way with respect to the hour angle $h \equiv t-\alpha$. In the following, it will be useful 
to use the fact that once integrated over $\alpha$ in the field of view available at each 
time $t$, the resulting function $\overline{\omega}$ depends only on the declination:
\begin{equation}
\label{eqn:omegad}
\overline{\omega}(\delta) \equiv \int^{t+\pi}_{t-\pi} \hspace{-0.4cm}d\alpha\,\omega(t-\alpha,\delta)
=2\int^{t}_{t-\pi} \hspace{-0.4cm}d\alpha\,\omega(t-\alpha,\delta) 
=2\int_{0}^\pi  \hspace{-0.2cm} dh\,\omega(h,\delta).
\end{equation}

In real experiments, the observed counting rates $I_E^{obs}(t)$ and $I_W^{obs}(t)$ may be 
modulated by small instrumental and atmospheric effects
that influence measurement conditions during the data taking. Weather effects on EAS development 
actually modulate the estimation of the energy and thus the counting rate due to the steep energy
spectrum, but they can formally be also treated as modulating in time the exposure of the
experiment. For instance, if we adopt a simple modulation of the form
$\omega(t-\alpha,\delta)(1+\eta(t))$, with
$\eta(t)$ the associated variation of the exposure, depending only
on the experimental conditions at local sidereal time $t$ and not on the direction
$(\alpha,\delta$), the observed counting rates can thus be expressed as:
\begin{eqnarray}
\label{eqn:rates_obs}
I_E^{obs}(t)&\simeq&I_E^{true}(t)+A\int_{\delta_{min}}^{\delta_{max}} \hspace{-0.6cm}d\delta\,\cos\delta\,\int_t^{t+\pi} \hspace{-0.4cm} d\alpha\,\omega(t-\alpha,\delta)\,\Phi_0\,\eta(t),\\
I_W^{obs}(t)&\simeq&I_W^{true}(t)+A\int_{\delta_{min}}^{\delta_{max}} \hspace{-0.6cm}d\delta\,\cos\delta\,\int^t_{t-\pi} \hspace{-0.2cm}d\alpha\,\omega(t-\alpha,\delta)\,\Phi_0\,\eta(t),
\end{eqnarray}
where we neglected second order terms proportional to $\Phi_1 \eta(t)$. Since the exposure and 
the associated variations are identical in the East and West, it is straightforward to see 
that through the subtraction $I_E^{obs}(t)-I_W^{obs}(t)$, the terms proportional to $\eta(t)$ cancel, leading to:

\begin{eqnarray}
\label{eqn:rates_h_2}
I_E^{obs}(t)-I_W^{obs}(t)&=&A \int_{\delta_{min}}^{\delta_{max}} \hspace{-0.8cm}d\delta\,\cos\delta\bigg[\int_t^{t+\pi} \hspace{-0.4cm}d\alpha\,\omega(t-\alpha,\delta)\,\Phi(\alpha,\delta)
-\int^t_{t-\pi} \hspace{-0.2cm}d\alpha\,\omega(t-\alpha,\delta)\,\Phi(\alpha,\delta)\bigg] \\
&\simeq& I_E^{true}(t)-I_W^{true}(t) \nonumber,
\end{eqnarray}
Assuming that the small angular 
scale variations of the flux $\Phi(\alpha,\delta)$ are small, we can approximate the 
integrations by considering the flux $\Phi(\alpha,\delta)$ as constant in each sector and 
evaluated at the right ascension of the mean exposed direction at each time 
$\alpha=t\pm\left<h\right>$, where $\left<h\right>$ is the mean hour angle, 
$\langle h \rangle = \int_{\delta_{min}}^{\delta_{max}}d\delta\,\cos\delta \int^t_{t-\pi}d\alpha\,\omega(t-\alpha,\delta)(t-\alpha)$. Then, 
\begin{equation}
\label{eqn:rates_h_3}
I_E^{obs}(t)-I_W^{obs}(t)\simeq A
\int_{\delta_{min}}^{\delta_{max}}  \hspace{-0.6cm}d\delta\,\cos\delta\bigg[\Phi(t+\left<h\right>,\delta)-\Phi(t-\left<h\right>,\delta)\bigg]\int^t_{t-\pi} \hspace{-0.2cm}d\alpha\,\omega(t-\alpha,\delta).
\end{equation}
The additional simplification is to consider the mean hour angle small enough with 
respect to the scale of variation of $\Phi$ so that one can use the linearised expressions  
$\Phi(t\pm \left<h\right>,\delta)\simeq\Phi(t,\delta)\pm \left<h\right>\partial{\Phi(t,\delta)}/\partial{t}.$   
Under these crude simplifications, the difference in the East and West counting rates can be 
approximated by:
\begin{eqnarray}
\label{eqn:rates_h_4}
I_E^{obs}(t)-I_W^{obs}(t)&\simeq&A \int_{\delta_{min}}^{\delta_{max}}  \hspace{-0.6cm}d\delta\,\cos\delta\bigg[2\left<h\right>\frac{\partial{\Phi(t,\delta)}}{\partial{t}}\bigg] \frac{\overline{\omega}(\delta)}{2} \nonumber \\
&\simeq&\left<h\right>\frac{d}{dt}\bigg[A\int_{\delta_{min}}^{\delta_{max}}  \hspace{-0.6cm}d\delta\,\cos\delta\,\Phi(t,\delta)\,\overline{\omega}(\delta)\bigg].
\end{eqnarray}
The expression between the brackets represents the true total counting 
rate, $I_{tot}^ {true}(t)=I_E^{true}(t)+I_W^{true}(t)$. Hence, we are led to the following 
relationship: 
\begin{eqnarray}
\label{eqn:rates_h_6}
I_E^{obs}(t)-I_W^{obs}(t) &\simeq&\left<h\right>\frac{dI_{tot}^ {true}(t)}{dt},
\end{eqnarray}
showing that the East-West counting rate difference, which is an observable quantity, is 
proportional to the derivative of the true intensity of cosmic rays, the coefficient of 
proportionality  being approximately the mean hour angle, which is also a measurable quantity.  

To derive the classical result, it was necessary to perform some crude simplifications to get at
Eqn.~\ref{eqn:rates_h_4} from Eqn.~\ref{eqn:rates_h_2}. These approximations are expected to 
become less accurate as the zenithal range of the experiment is increased. However, we will see 
in the next section that the structure of the relationship still holds (with a different 
proportionality factor) in the general case. 

\section{A more rigorous description of the East-West method} \label{section:ew}

We repeat here the previous calculations but without making the same simplifications as
in the classical implementation. The aim is to check whether the relationship between 
the differential counting rate $I_E^{obs}-I_W^{obs}$ and the derivative of the total 
counting rate $dI_{tot}^{true}/dt$ still holds or not. 

It turns out that the calculation 
is easier when performed in local coordinates. The counting rates $I_E^{obs}(t)$ 
and $I_W^{obs}(t)$ at local sidereal time $t$ for the two halves of the sky can be 
computed from the cosmic ray flux $\Phi$ expressed in local coordinates 
$(\theta,\phi)$ as:
\begin{eqnarray}
\label{eqn:rates}
I_E^{obs}(t)&=&A \int_{-\pi/2}^{\pi/2}\mathrm{d}\phi\,\int_{0}^{\theta_{\mathrm{max}}} \mathrm{d}\theta\,\sin{\theta}\cos{\theta}\,\epsilon(\theta,t)\,\Phi(\theta,\phi,t),\nonumber \\
I_W^{obs}(t)&=&A \int_{\pi/2}^{3\pi/2}\mathrm{d}\phi\,\int_{0}^{\theta_{\mathrm{max}}} \mathrm{d}\theta\,\sin{\theta}\cos{\theta}\,\epsilon(\theta,t)\,\Phi(\theta,\phi,t),
\end{eqnarray}
where $A\cos{\theta}$ is the effective area of the experiment at an angle of 
incidence $\theta$ and $\epsilon(\theta,t)$ is the detection  
efficiency function which includes the time-dependent spurious
effects. Here, we adopt the convention that the  
azimuth angle is defined relative to the East direction, measured
counterclockwise. To guarantee that the Eastern and Western  
sectors are equivalent in terms of counting rates, any dependence of
$\epsilon$ in azimuth $\phi$ needs to be symmetrical.  
For simplicity, we assume hereafter a uniform detection efficiency in
azimuth; but similar conclusions still hold as long as  
the symmetry between the sectors is respected, which is a reasonable
assumption in practice. It is also reasonable to assume  
that the relative amplitude $\eta$ of the temporal variations of the exposure is
small, and that those variations decouple from  
the zenith angle dependent ones:
\begin{equation}
\epsilon(\theta,t)=\epsilon_1(\theta)(1+\eta(t)).
\end{equation}

To get an explicit expression for the cosmic ray flux $\Phi$
in local coordinates, we 
start from the parameterisation $\Phi(\alpha,\delta)$ in terms of the
equatorial coordinates. The most basic approach to probe  
a large scale variation is to describe the flux by a combination of an isotropic
component and a dipolar component: 
\begin{equation}
\Phi(\alpha,\delta)=\Phi_0\cdot\bigg(1+\hat{u}(\alpha,\delta)\cdot\vec{D}\bigg),\hspace{1cm}\vec{D}=D\cdot \hat{u}(\alpha_d,\delta_d),
\end{equation}
with $\vec{D}$ being the dipole vector defined by its magnitude $D$
and its orientation $(\alpha_d,\delta_d)$ 
($\hat{u}(\alpha,\delta)$ denotes the unit vector in the direction 
$(\alpha,\delta)$).  
As the conversion between equatorial and horizontal coordinates is
time-dependent, the flux from  
a specific viewing direction ($\theta,\phi$) expressed in local
zenithal coordinates at any given location on the Earth turns out 
to be time-dependent. Adopting local coordinates with the $z$ axis in 
the zenithal direction, the $x$ axis towards the East and the $y$ axis 
towards the North, the dipolar vector $\vec{D}(t)$ is written as
\begin{eqnarray}
\vec{D}(t)=\left( \begin{array}{lll} 
D_x(t)\\
D_y(t)\\
D_z(t) \end{array} \right)
=D\cdot \left( \begin{array}{lll} 
\sin{\theta_d(t)}\cos{\phi_d(t)}\\
\sin{\theta_d(t)}\sin{\phi_d(t)}\\
\,\,\,\,\,\,\,\,\,\,\,\cos{\theta_d(t)} \end{array} \right).
\end{eqnarray}
Transforming from local to equatorial coordinates, this can be expressed as:
\begin{eqnarray}
\vec{D}(t)=D\cdot \left( \begin{array}{lll} 
\,\,\,\,\,\,\,\,\,\,\,\,\,\,\,\,\,-\cos{\delta_d}\sin{h_d(t)}\\
\cos{\ell}\sin{\delta_d}-\sin{\ell}\cos{\delta_d}\cos{h_d(t)}\\
\sin{\ell}\sin{\delta_d}+\cos{\ell}\cos{\delta_d}\cos{h_d(t)} \end{array} \right),
\end{eqnarray}
with $h_d(t)=t-\alpha_d$ the hour angle of the dipole at the location of the observation point and at time $t$, and $\ell$
the Earth latitude of the observation point. Inserting this expression 
in $\Phi(\theta,\phi,t)$  and carrying out 
the integrations over $\theta$ and $\phi$ in Eqn.~\ref{eqn:rates} leads to:
\begin{eqnarray}
I_E^{obs}(t)&=&A\Phi_0\cdot ( \pi g_{11}(t)+2D_x(t)g_{12}(t)+\pi D_z(t)g_{21}(t)),\\
I_W^{obs}(t)&=&A\Phi_0\cdot ( \pi g_{11}(t)-2D_x(t)g_{12}(t)+\pi D_z(t)g_{21}(t)),
\end{eqnarray}
where the $g_{ij}(t)$ coefficients are defined by: 
\begin{eqnarray}
g_{ij}(t)&=&(1+\eta(t))\int_{0}^{\theta_{\mathrm{max}}}\,d\theta\,\epsilon_1(\theta)\,\cos^i{\theta}\sin^j{\theta}\\
&=&(1+\eta(t))f_{ij}.
\end{eqnarray}
Once normalised by $f_{11}$, each $f_{ij}$ coefficient can be estimated directly from 
any data set through the empirical averages of $\cos^{i-1}\theta\cdot\sin^{j-1}\theta$:
\begin{equation}
\frac{f_{ij}}{f_{11}}=\left<\cos^{i-1}\theta\cdot\sin^{j-1}\theta\right>.
\end{equation}
As long as $\theta_{\mathrm{max}}$ is smaller than $\simeq 70^\circ$, the zenithal distribution
of CRs is only marginally affected by any large scale anisotropy patterns, in such a way that such 
empirical estimations, rigorous in case of isotropy, are still accurate enough even in the 
presence of genuine anisotropies. Then, by calculating as previously 
$I_E^{obs}(t)-I_W^{obs}(t)$ and $dI_{tot}^{true}/dt$, and by neglecting second order terms
proportional to $\eta D$, we find now that: 
\begin{eqnarray}
I_E^{obs}(t)-I_W^{obs}(t)&\simeq&-4A\Phi_0Df_{12}\cos\delta_d\sin{h_d(t)},\\
\label{eqn:didt}\frac{dI_{tot}^{true}(t)}{dt}&\simeq&-2\pi A\Phi_0Df_{21}\cos\ell\cos\delta_d\sin{h_d(t)}.
\end{eqnarray}
We then finally obtain:
\begin{equation}
\label{eqn:ew}
I_E^{obs}(t)-I_W^{obs}(t)\simeq\frac{2}{\pi\cos\ell} \frac{\left<\sin\theta\right>}{\left<\cos\theta\right>} \frac{dI_{tot}^{true}(t)}{dt}.
\end{equation}
Consequently, the relationship between the East-West counting rate
difference and the differential total counting rate is at first order 
similar to that discussed in the previous section, except that the 
proportionality factor is now given by  $h_\star \equiv 2\left<\sin\theta\right>/\pi\cos\ell\left<\cos\theta\right>$,
which can also be calculated from the measured zenith angles of the events. 

\subsection{Comparison with the classical implementation} 

\begin{figure}[t]
\begin{center}
{\includegraphics[width=10.5cm]{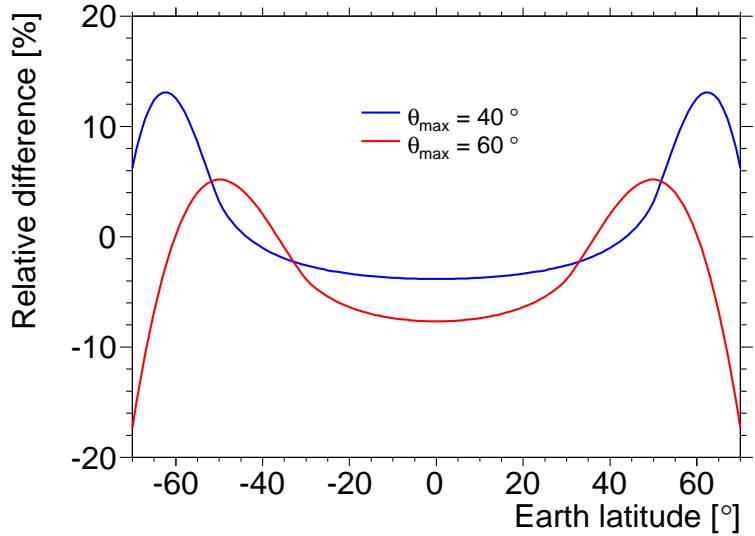}}
\end{center}\caption{Relative differences between the mean hour angle and the factor $2\left<\sin\theta\right>/\pi\cos\ell\left<\cos\theta\right>$
as a function of the Earth latitude, for different conditions of maximal zenith angle available in the field of view.} 
\label{fig:comp_fact}
\end{figure}

Despite the crude approximations outlined in Section \ref{subsec:class_implementation}, it turns
out that the factor $h_\star$ 
is not very different from the mean hour angle as long as the Earth
latitude of the site of the experiment is smaller than 50$^\circ$ 
in absolute value. This is illustrated in Fig.~\ref{fig:comp_fact},
where the relative difference 
$(\left<h\right>-h_\star)/h_\star$ is shown as a function of the
Earth latitude of the site, using 
$\epsilon(\theta,t)=1$ and two different values of $\theta_{max}$. It
can be seen that these relative differences  
become larger than 10$\%$ only for Earth latitudes
$\ell\geq 50^\circ$. The results would not 
be very different had we considered other examples of $\epsilon(\theta,t)$.

\section{First Harmonic Analysis - Estimation of the dipole equatorial component} \label{section:fha}

\subsection {First Harmonic Analysis of $I_E-I_W$}
\label{subsec:FirstHarmAnalysis}

To probe a dipolar structure of the CR arrival direction distribution,
Eqn. \ref{eqn:ew} is an ideal 
starting point to estimate the dipolar modulation of $dI_{tot}^{true}/dt$
parameterised through the amplitude $r$ and 
the phase $\varphi$: 
\begin{equation}
\frac{dI_{tot}^{true}(t)}{dt}=r\cos(t-\varphi).
\end{equation} 

From a set of $N$ arrival times from events coming from either the
Eastern or the Western directions, $r$ and $\varphi$ 
can be estimated by applying to the arrival times of the events the
standard first harmonic analysis~\cite{Linsley}  
slightly modified to account for the subtraction of the Western
sector to the Eastern one. The Fourier coefficients 
$a$ and $b$ are thus defined by:
\begin{eqnarray}
a &=& \frac{2}{N}\sum_{i=1}^{N} \cos{(t_i+\zeta_i)}, \\
b &=& \frac{2}{N}\sum_{i=1}^{N} \sin{(t_i+\zeta_i)},
\end{eqnarray}
where $\zeta_i$ equals 0 if the event is coming from the East or $\pi$
if coming from the West\footnote{We are grateful to Paul Sommers for suggesting
this simple way of accounting for the difference between the contributions from
the East and West sectors.}. The amplitude and 
phase estimates $(\hat{r},\hat{\varphi})$ of $dI_{tot}^{true}/dt$ are then
obtained through:
\begin{equation}
\hat{r}=\frac{\pi\cos\ell}{2}\frac{\left<\cos\theta\right>}{\left<\sin\theta\right>}\sqrt{a^{2}+b^{2}}\\
\hspace{1 cm} \mbox{and}\\
\hspace{1 cm} \hat{\varphi}=\arctan\bigg(\frac{b}{a}\bigg).
\end{equation}
By integrating $dI/dt$, the amplitude and phase estimates $(\hat{r}_I,\hat{\varphi}_I)$ of the
intensity $I(t)^{true}_{tot}$ itself are obtained:
\begin{equation}
\hat{r}_I=\frac{N}{2\pi}\hat{r}\\
\hspace{1 cm} \mbox{and}\\
\hspace{1 cm} \hat{\varphi}_I=\hat{\varphi}+\frac{\pi}{2}.
\end{equation}

\subsection{Estimation of the dipole equatorial component}

In case of the standard Rayleigh analysis in right-ascension, the first harmonic 
amplitude $r$ is related to the dipole amplitude $D$ through~\cite{aublin}:
\begin{equation} \label{eqn:ra}
r_{RA}=\left|\frac{\left<\cos\delta\right>D_\perp}{1+\left<\sin\delta\right>D_\parallel}\right|,
\end{equation}
where $D_\parallel=D\sin\delta_d$ denotes the component of the dipole along the Earth 
rotation axis while $D_\perp=D\cos\delta_d$ is the component in the equatorial plane. 
The first harmonic amplitude thus depends on the declination of the dipole in such a 
way that it vanishes for $\delta_d=\pm\pi/2$. This is obvious, as the modulation of the 
flux does not depend on the right ascension in such a case. On the other hand, the power 
of the method is the largest when the dipole is oriented in the equatorial plane. In this latter 
case the first harmonic amplitude becomes $r_{RA}=D_\perp\left<cos\delta\right>$ and the 
sensitivity of an experiment to the true value of $D_\perp$ depends then on 
$\left<\cos\delta\right>$, which is a function of the Earth latitude $\ell$ of the 
experiment and its detection efficiency in the zenithal range considered. 

Similarly, the first harmonic amplitude reconstructed by the East-West method is not 
directly the dipole amplitude. The additional step consists in relating $dI_{tot}^{true}/dt$ 
and the dipolar parameters through measurable quantities. Remembering that the observed number of 
events $N$ is equal to 
$\int_{2\pi}dt(I_E^{obs}+I_W^{obs})=2\pi A\Phi_0\int_{2\pi}dt(g_{11}(t)+D_z(t)g_{21}(t))$
and that 
$D_z(t)=D\cos{\theta_d(t)}=D\left(\sin{\ell}\sin{\delta_d}+\cos{\ell}\cos{\delta_d}\cos{h_d(t)}\right)$,  
and neglecting all second order terms in $\eta D_\perp$ and $\eta D_\parallel$, Eqn.~\ref{eqn:didt} can be rewritten as:
\begin{eqnarray}
\frac{dI_{tot}^{true}(t)}{dt}&\simeq&-\frac{N}{2\pi}\frac{D_\perp\left<\cos\theta\right>\cos\ell\sin{h_d(t)}}{1+\frac{1}{2\pi}\int dt\,\eta(t)+D_\parallel\left<\cos\theta\right>\sin\ell}
\end{eqnarray}
This expression describes how a genuine dipole with amplitude $D$ and 
pointing to a declination $\delta_d$ modulates in time  
the intensity of CRs observed by a ground experiment located at an
Earth latitude $\ell$. The amplitude of the time variation is suppressed
if the dipole is approximately aligned with the Earth rotation axis,
while the modulation is maximal if $D_\parallel$ is small compared
to $D_\perp$. In this latter case the differential counting rate can
thus be expressed as a function of the equatorial  
component of the dipole in a straightforward way:
\begin{equation}
\label{eqn:eqdip}
I_E^{obs}(t)-I_W^{obs}(t)\simeq-\frac{N}{2\pi}\frac{2\left<\sin\theta\right>}{\pi}D_\perp\sin{h_d(t)},
\end{equation}
where we have neglected the second order terms proportional to $\eta D_\perp$ and $D_\parallel D_\perp$.

The first harmonic amplitude and phase estimates $(\hat{D}_\perp,\hat{\alpha}_d)$ 
are thus obtained through: 
\begin{equation}
\label{eqn:ampl_ph}
\hat{D}_\perp=\frac{\hat{r}}{\cos\ell\left<\cos\theta\right>}=\frac{\pi}{2\left<\sin\theta\right>}\sqrt{a^2+b^2}\\
\hspace{1 cm} \mbox{and}\\
\hspace{1 cm} \hat{\alpha}_d=\hat{\varphi}+\frac{\pi}{2}.
\end{equation}
It is worth noting that, from the transformation of coordinates
relation $\cos\delta\sin{h}=-\sin\theta\cos\varphi$,  
the factor $2\left<\sin\theta\right>/\pi$ can be expressed as well in
terms of $\delta$ and $h$ as: 
\begin{equation}
\label{eqn:relation}
\frac{2\left<\sin\theta\right>}{\pi}=\left<\cos\delta\sin{h}\right>,
\end{equation}
where the r.h.s. average is performed by integrating over the eastern and western 
quadrants.

By comparing Eqn.~\ref{eqn:ampl_ph} to Eqn.~\ref{eqn:ra} and using Eqn.~\ref{eqn:relation}, 
it can be seen that the use of 
the local sidereal time in the modulation search (instead of the right ascension as in the case
of the standard Rayleigh analysis), combined to the East-West subtraction, leads to a loss 
of sensitivity by a factor $\left<\cos\delta\right>/\left<\cos\delta\sin{h}\right>$ 
(which is typically about two, depending on the experimental conditions) 
with respect to the performances of the standard Rayleigh analysis. However, this 
method has the benefit of avoiding the need to implement any corrections of the total counting 
rates for instrumental and atmospheric effects. Moreover, in some cases those
corrections cannot be computed reliably, for instance when they are due to
the energy dependence of the trigger efficiency, which can also depend on
the unknown composition of the primary CRs. In these cases, only the East-West
method can be implemented reliably.

\section {Simulations} \label{section:mc}

In this section, we check the behavior of the method through
simulations reproducing realistic conditions of a ground 
experiment subject to artificial modulations at both the diurnal and the
seasonal time scales. For this purpose, we consider three typical cases of interest 
for this kind of studies:
\begin{itemize}
    \item isotropy;
    \item genuine sidereal modulation;
    \item genuine solar modulation.
\end{itemize}
An additional underlying distribution has been considered to study the impact 
of a broken East-West symmetry:
\begin{itemize}
    \item isotropy with a broken East-West symmetry.
\end{itemize}
In order to test extreme situations we also add the spurious effects to all of these distributions.

For definiteness  we consider an experiment located  
at the same Earth latitude $\ell=-$35.25$^\circ$ as the Pierre Auger
Observatory~\cite{augerdet} and sensitive up to a maximal zenith angle
of 60$^\circ$, with the following  
energy independent detection efficiency function:
\begin{equation}
\epsilon(\theta,t)=\frac{1}{1+\exp((\theta-\theta_{ns})/\sigma_{ns})} \cdot g(t), 
\end{equation}
where $\theta_{ns}=50^\circ$ and $\sigma_{ns}=5^\circ$. Here, $g(t)$ stands for
the spurious modulation, which may differ from unity due to various reasons. 
  For instance, the
changes of atmospheric conditions affect the energy  
estimate of the showers~\cite{Auger_weather}. When such effects are
not accounted for, they induce a modulation of the rate of 
events above any given energy threshold. Hence, we choose the function $g(t)$ to be of the generic  
form~\cite{Farley&Storey}:
 \begin{equation}
g(t)=\frac{1+\eta_y \cos(2\pi(t-t^0_y)/T_y)+\eta_d\bigg[1+\eta_\star \cos(2\pi(t-t^0_y)/T_y)\bigg] \cos(2\pi(t-t^0_d)/T_d)}{1+\eta_y+\eta_d(1+\eta_\star)},
\end{equation}
where $t$ is here expressed in terms of solar time. The cosine
proportional to $\eta_y$ describes the annual variation  
of the mean event rate, with phase $t^ 0_y$ and a period $T_y$ of
one year. The cosine proportional to $\eta_d$ describes  
the annual average of the solar diurnal modulation, with phase $t^0_d$ 
and a period $T_d$ of one day. Finally, $\eta_\star$  
stands for the variation of the diurnal amplitude along the year. This
last term, combining the diurnal modulation with an 
annual one, is responsible for the production of sidebands at both the
sidereal and the anti-sidereal frequencies,  
whose amplitudes are given by
$0.5\times\eta_d\times\eta_\star$~\cite{Farley&Storey}. To
illustrate the power of  
the East-West method, we choose an extremely high value of
$\eta_\star=90\%$ to guarantee the existence of significant 
sidebands, while we set $\eta_y=20\%$ and $\eta_d=15\%$.

\subsection{Case 1: isotropy with spurious effects}

\begin{figure}[htbp]
\begin{center}
\subfigure[]
{\includegraphics[width=8.5cm]{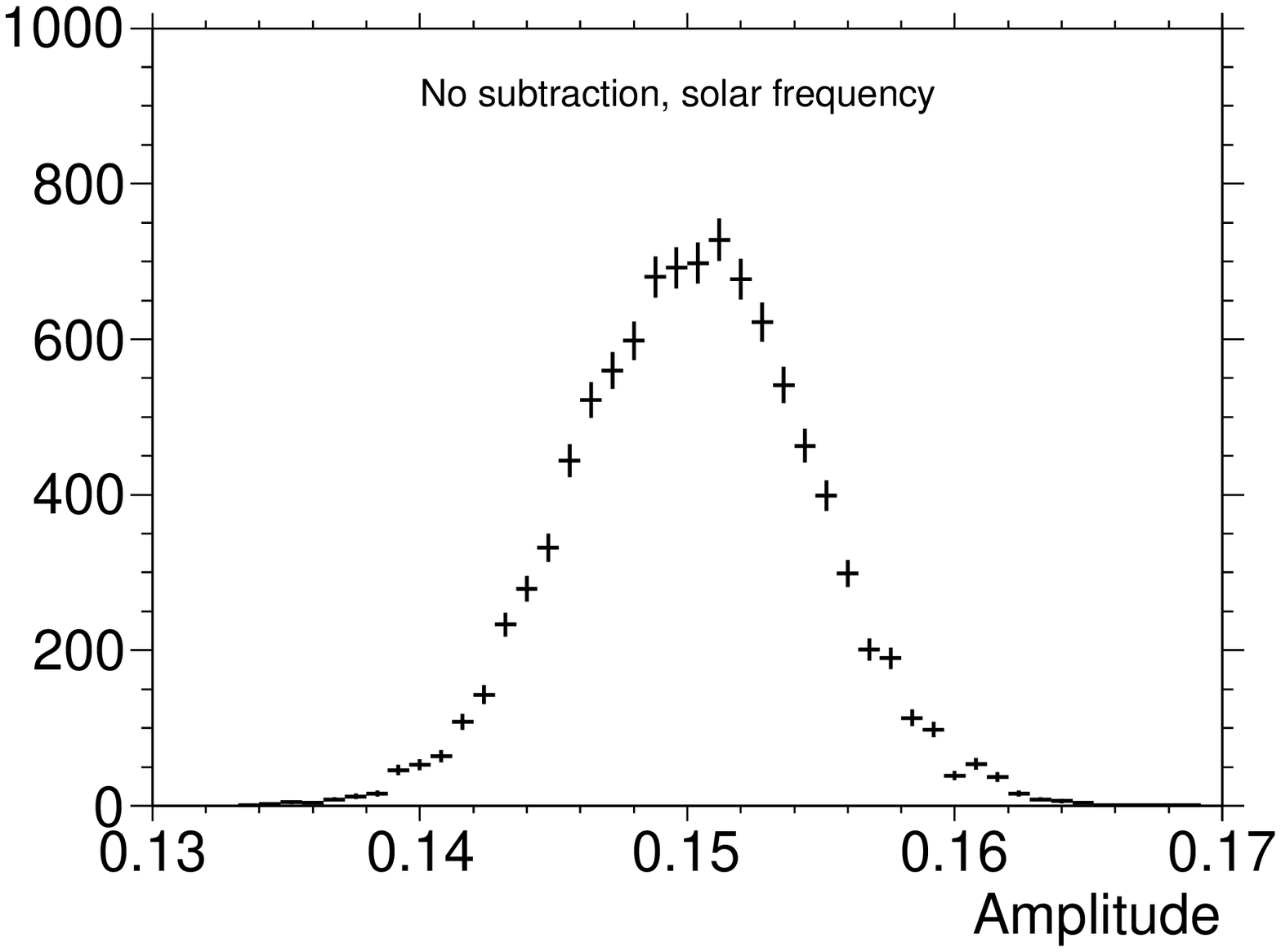}}
\subfigure[]
{\includegraphics[width=8.5cm]{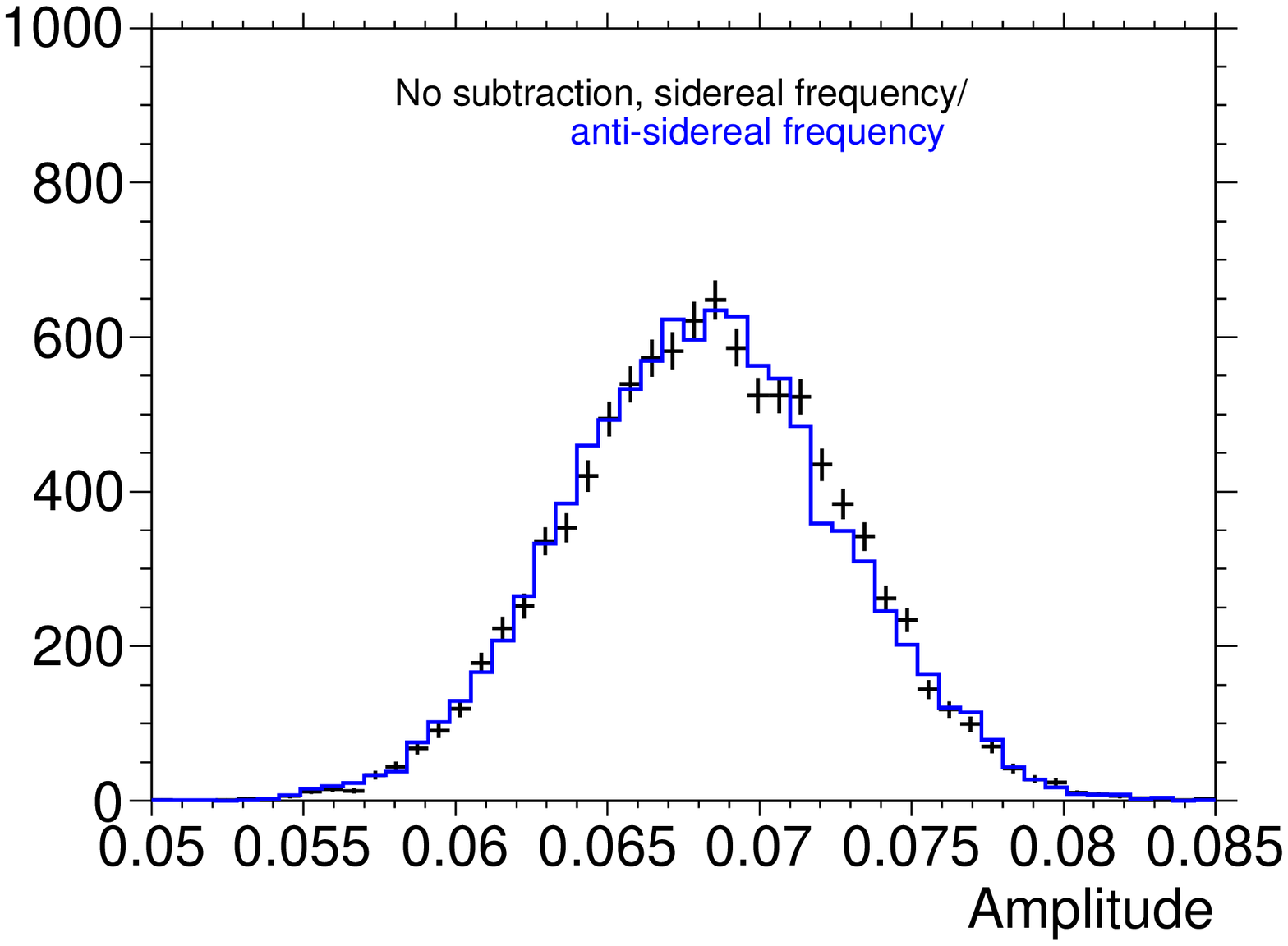}}
\subfigure[]
{\includegraphics[width=8.5cm]{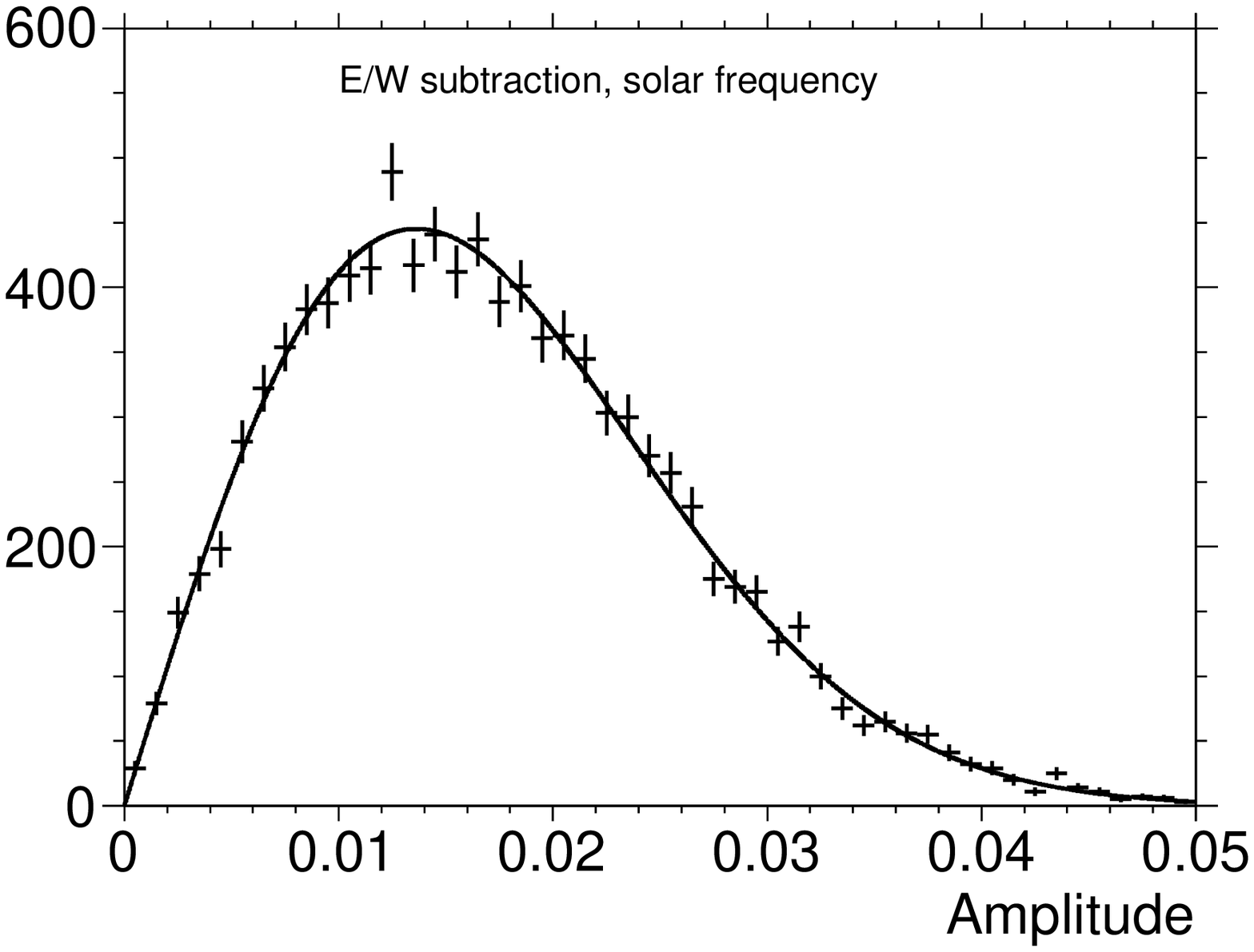}}
\subfigure[]
{\includegraphics[width=8.5cm]{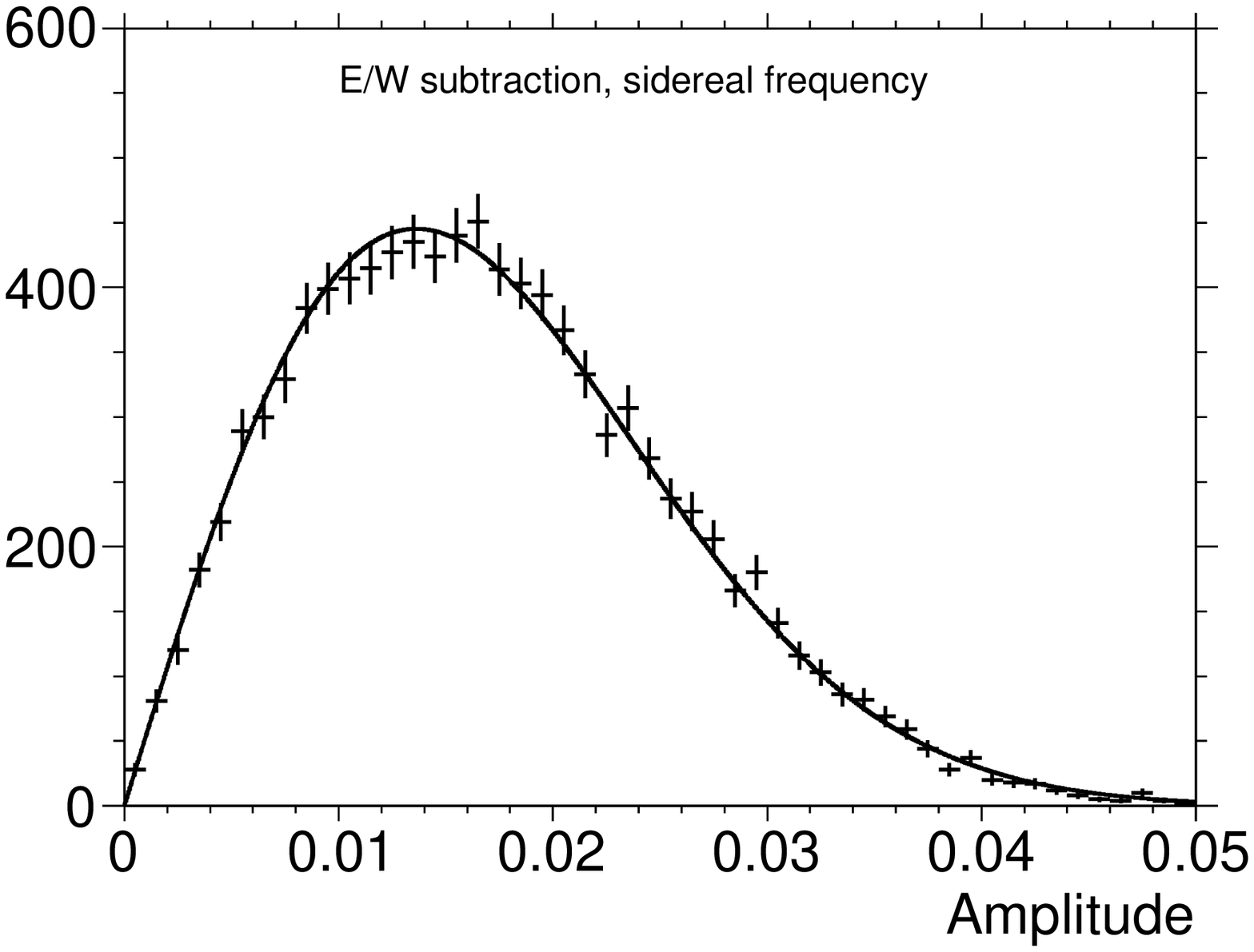}}
\subfigure[]
{\includegraphics[width=8.5cm]{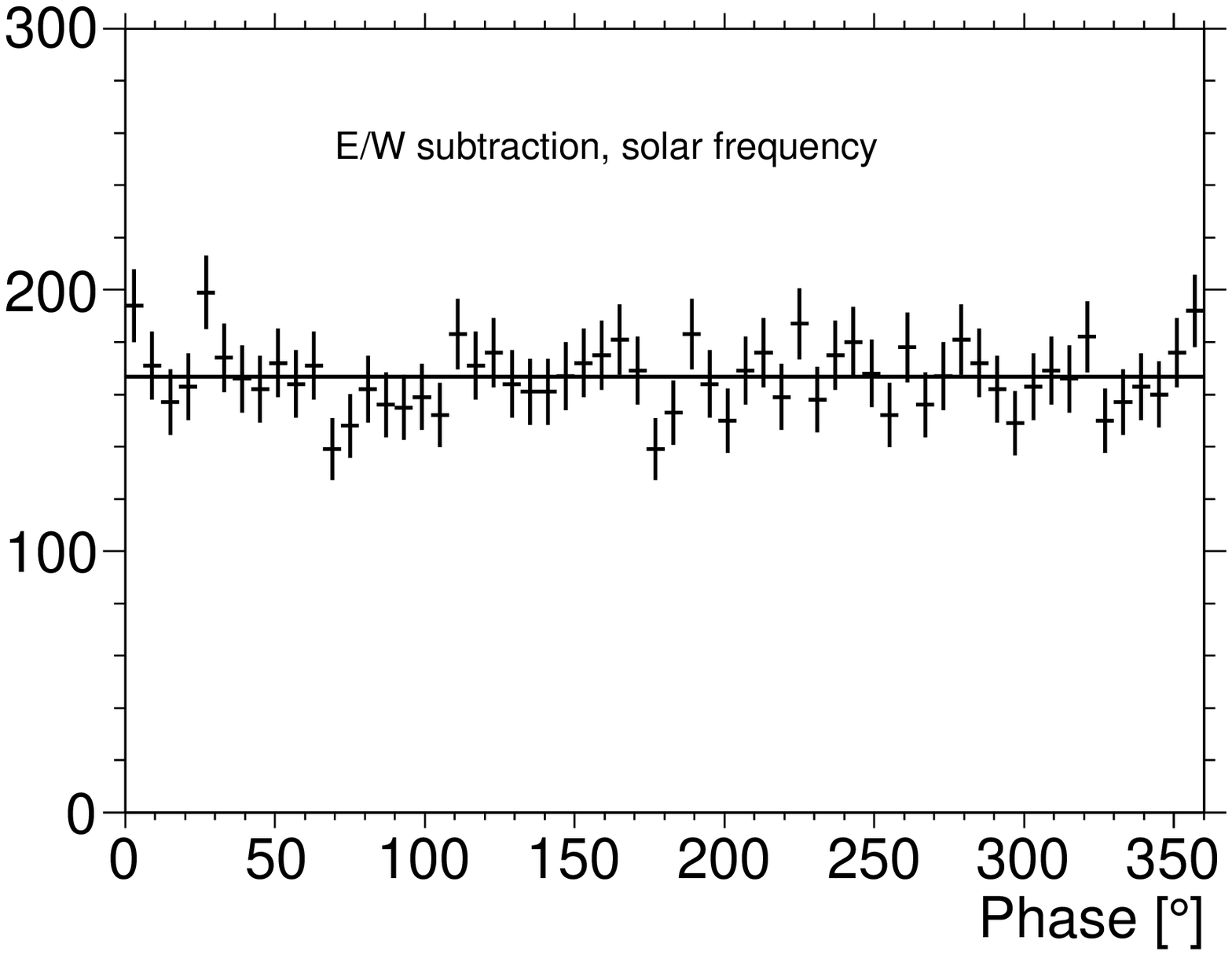}}
\subfigure[]
{\includegraphics[width=8.5cm]{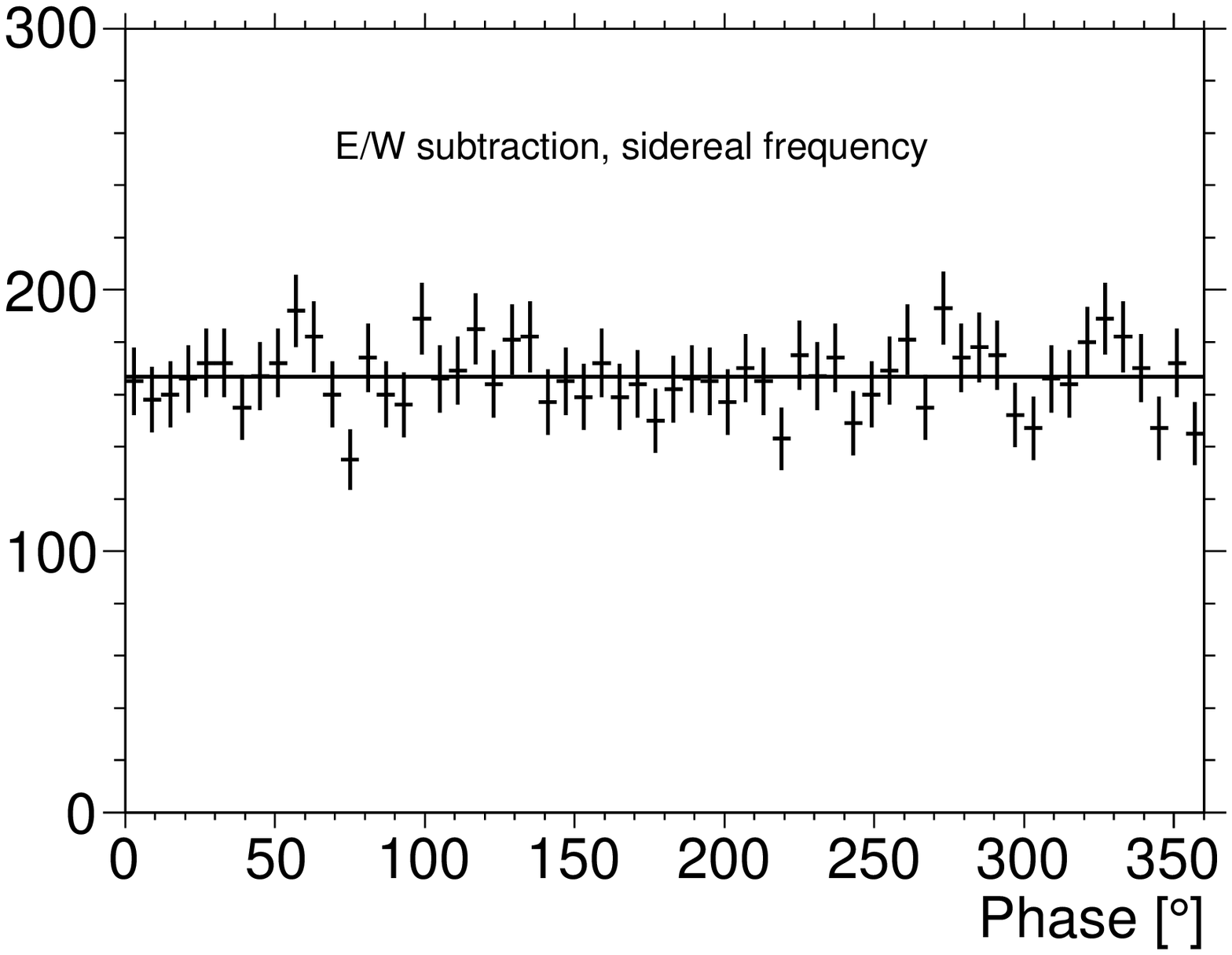}}
\end{center}\caption{Top: Distributions of the amplitudes of the first harmonic obtained with the 
standard Rayleigh analysis in the presence of experimental effects in solar time ($a$) and in 
both sidereal and anti-sidereal times ($b$). Middle and Bottom: Distributions of the amplitudes 
($c$ and $d$) and phases ($e$ and $f$) of the first harmonic obtained with the East-West 
analysis, compared with the isotropic expectations: the subtraction of the spurious effects holds 
perfectly.} 
\label{fig:case1}
\end{figure}

We first consider an isotropic distribution of CRs polluted by the
spurious effects described above, and  
analyse $10^4$ mock samples generated with a total number of events
$N=10^5$. The net results of the  
instrumental and atmospheric effects introduced by the function $g(t)$ are shown on top of 
Fig.~\ref{fig:case1}, evidencing that the net counting rate  
$I^{obs}_{tot}=I^{obs}_E+I^{obs}_W$ undergoes the expected modulation
of amplitude $\eta_d=15\%$ at the solar  
frequency and of $0.5\times\eta_d\times\eta_\star=6.75\%$ at both the
sidereal and the anti-sidereal frequencies. 
Applying the East-West analysis at both the solar and sidereal
frequencies, it can be seen that the reconstructed 
amplitudes $\hat{D}_\perp$ and phases $\hat{\alpha}_d$ are now
distributed according to the expected distributions:  
the Rayleigh one for the amplitude with parameter
$\sigma=\pi/2\left<\sin\theta\right>\sqrt{2/N}$, and the uniform one  
for the phase. Hence, in spite of the strong experimental effects, it turns
out that the East-West subtraction allows the
removal of possible biases in the estimate of both the amplitude and
phase in the case of an underlying isotropic distribution of CRs.

\subsection{Case 2: 5\% sidereal signal with spurious effects on the acceptance}
\label{subsec:case2}

\begin{figure}[t]
\begin{center}
\subfigure[]
{\includegraphics[width=8.5cm]{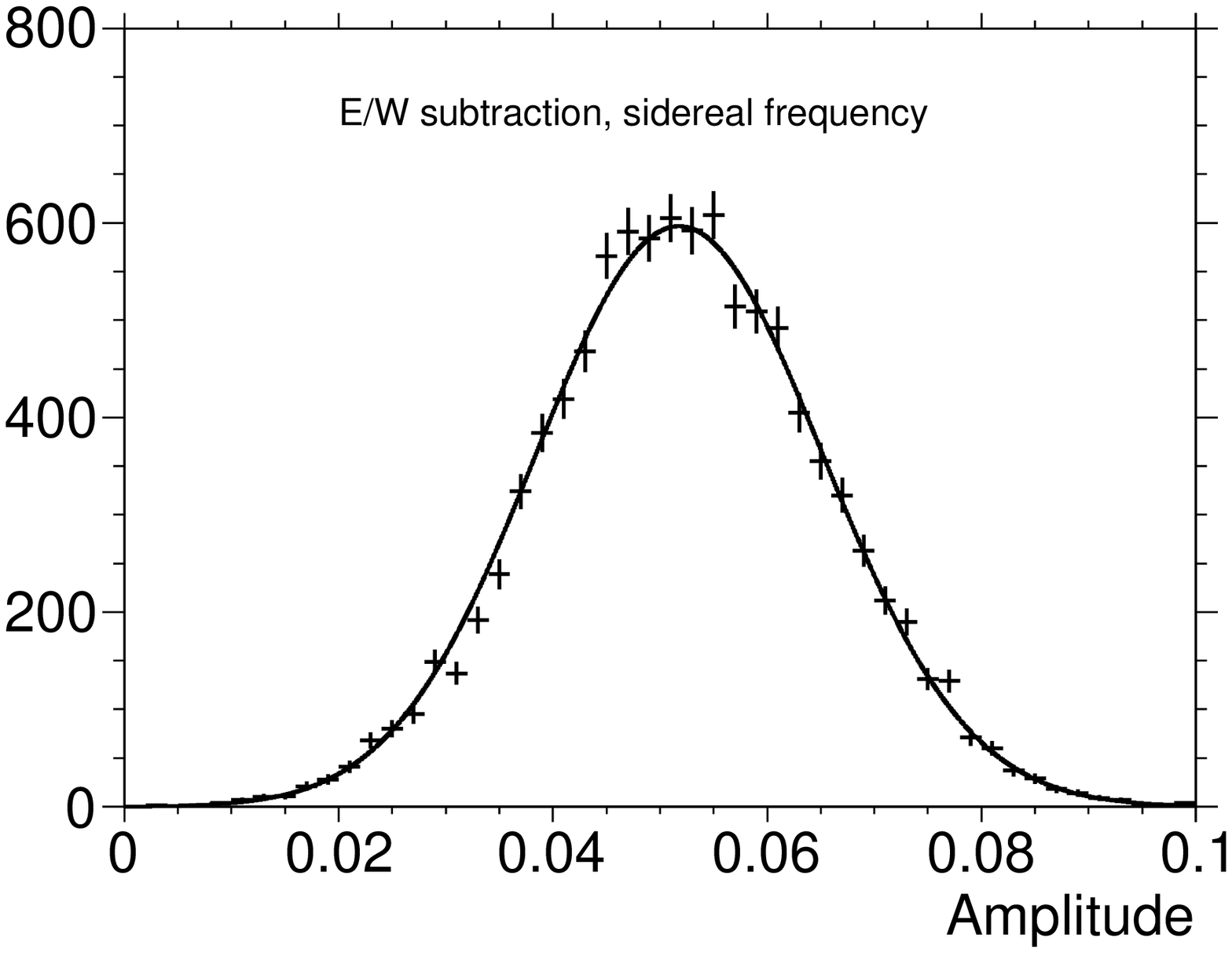}}
\subfigure[]
{\includegraphics[width=8.5cm]{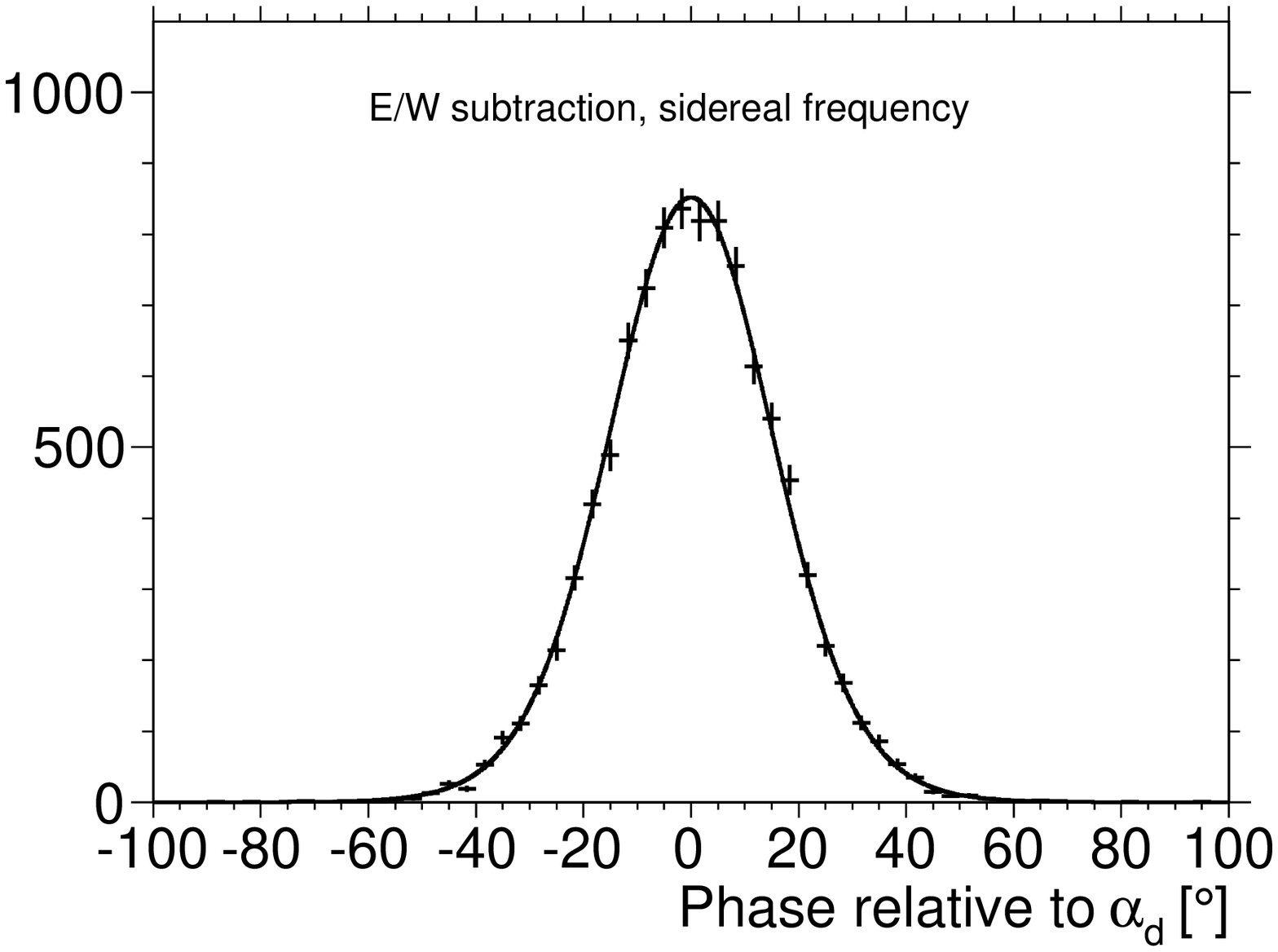}}
\end{center}\caption{Distributions of the amplitudes  ($a$) and phases
  ($b$) of the first harmonic in local sidereal time obtained with the
  East-West 
analysis in the presence of a genuine pattern of $5$\%. Both histograms
are in perfect agreement with respect to the expectations.}  
\label{fig:case2}
\end{figure}

To test now the accuracy of the method in the presence of both a genuine
signal at the sidereal frequency and spurious 
effects, a signal corresponding to  a dipolar anisotropy
of 5\% amplitude is introduced in the simulated samples. For
definiteness, we consider the dipole  oriented towards the 
equatorial plane. The reconstructed amplitudes are now expected to
follow a Rice distribution with parameters  
$\mu=5\%$ and $\sigma=\pi/2\left<\sin\theta\right>\sqrt{2/N}$:
\begin{equation}
p_1(\hat{D}_\perp)=\frac{\hat{D}_\perp}{\sigma^2} \exp{\bigg(-\frac{\hat{D}_\perp^2+\mu^2}{2\sigma^2}\bigg)}I_0\bigg(\frac{\hat{D}_\perp\mu}{\sigma^2}\bigg),
\end{equation}
while the reconstructed phases are expected to follow the distribution
described by Linsley in the 2$^{nd}$ alternative in~\cite{Linsley}:
\begin{equation}
p_2(\hat{\alpha}_d)=\frac{1}{2\pi}\exp{\bigg(-\frac{\mu^2}{2\sigma^2}}\bigg)+\frac{\mu\cos{(\hat{\alpha}_d-\alpha_d)}}{2\sqrt{2\pi}\sigma}\bigg(1+\mathrm{erf}{\bigg(\frac{\mu\cos{(\hat{\alpha}_d-\alpha_d)}}{\sqrt{2}\sigma}\bigg)}\bigg)\exp{\bigg(-\frac{\mu^2\sin^2{(\hat{\alpha}_d-\alpha_d)}}{2\sigma^2}\bigg)},
\end{equation}
The results of the simulations are shown in Fig.~\ref{fig:case2}, evidencing a perfect agreement with the expectations.

\subsection{Case 3: $0.1$\% solar signal with spurious effects}

\begin{figure}[t]
\begin{center}
\subfigure[]
{\includegraphics[width=8.5cm]{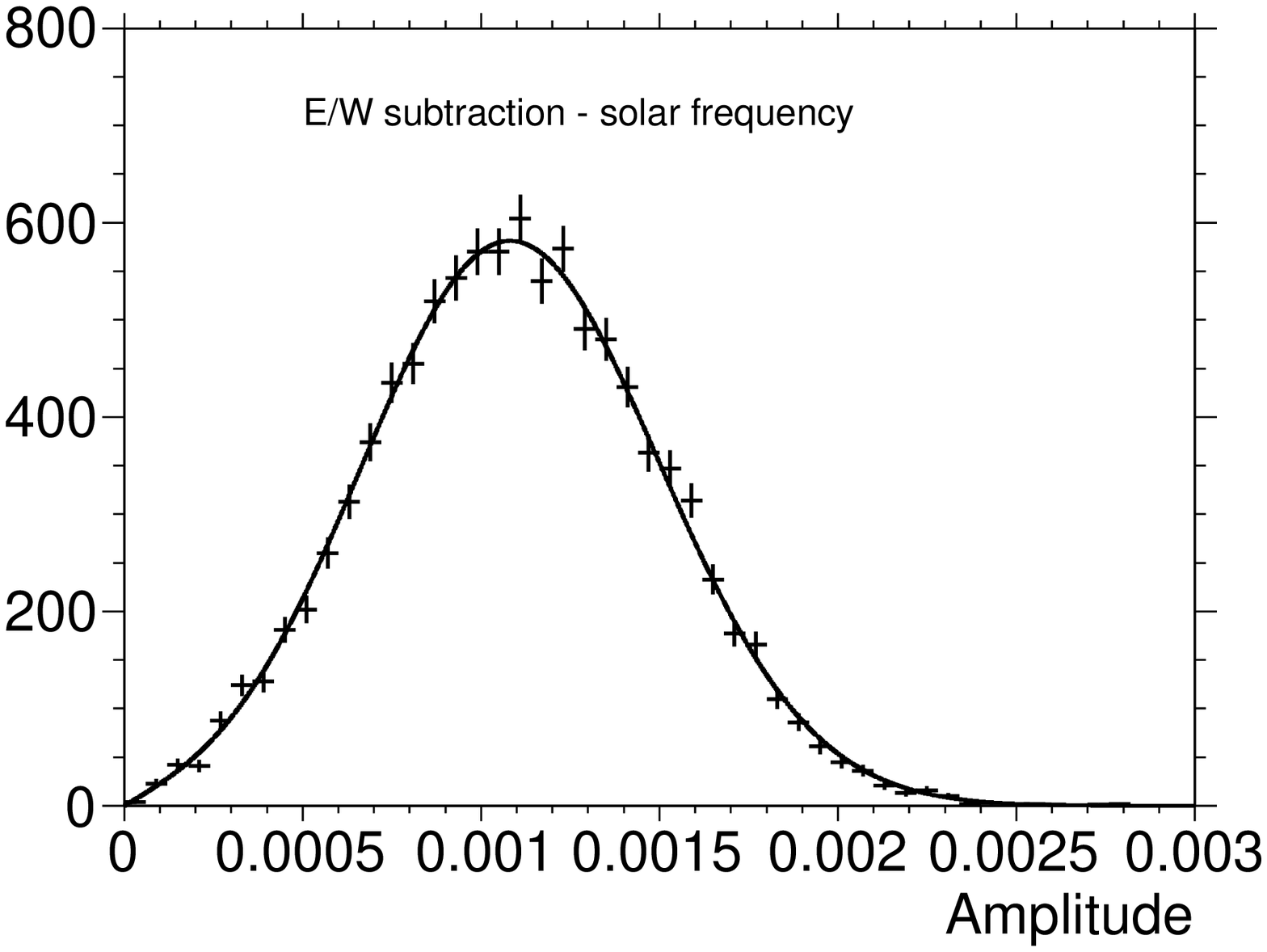}}
\subfigure[]
{\includegraphics[width=8.5cm]{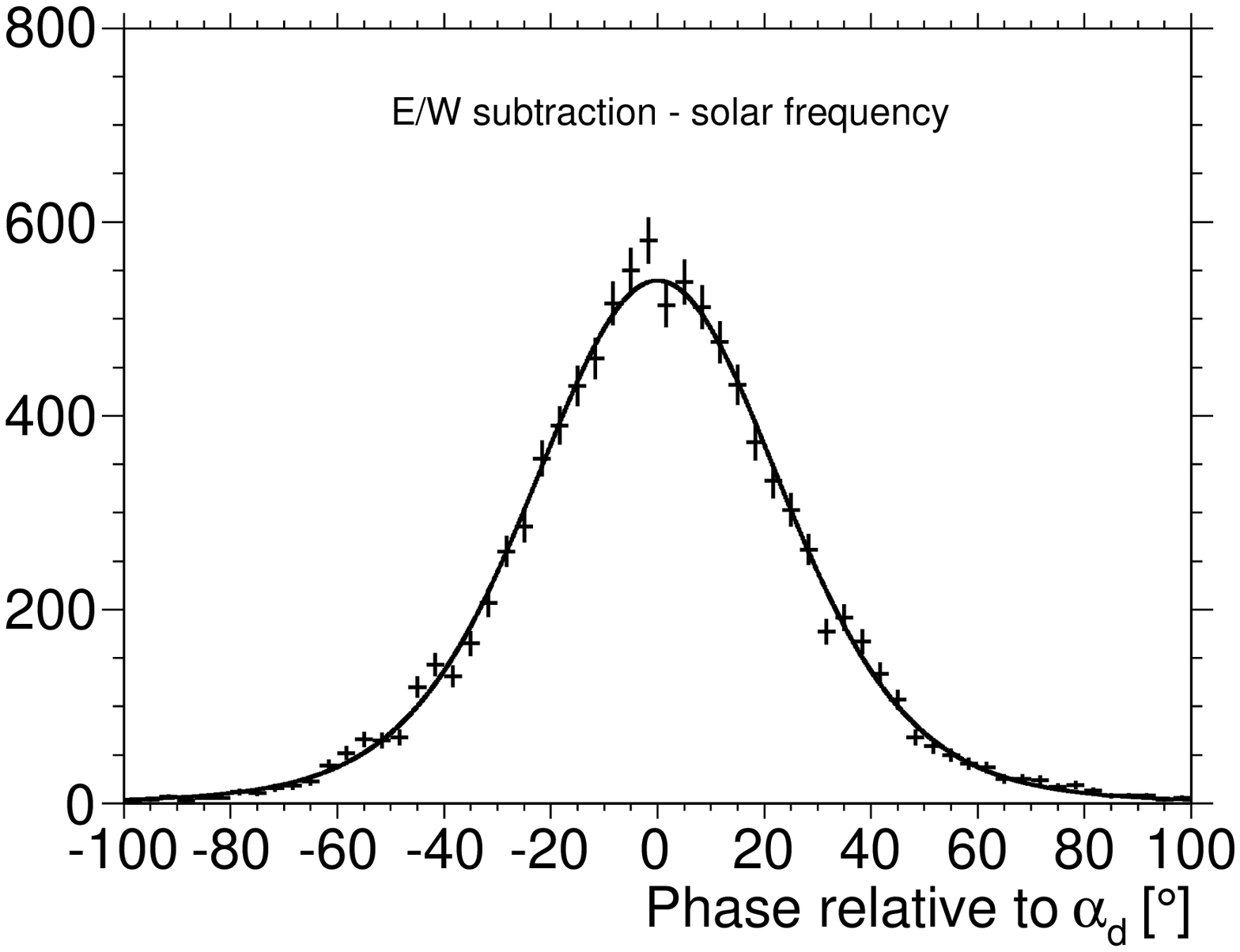}}
\end{center}\caption{Same as Fig.~\ref{fig:case2}, but generating a
  genuine pattern at the solar frequency.}  
\label{fig:case3}
\end{figure}

We repeat  here exactly the same exercise as above, but
generating a genuine dipole with an amplitude $0.1$\% at the  
\textit{solar} frequency, together with the spurious effects. This
kind of feature is expected  due to  
the motion of the terrestrial observer through the frame in which the
CR distribution is isotropic. It has been observed by several
experiments at low energies, where sufficient statistics has been
gathered. To probe    
such a low amplitude, the number of events has to be greatly increased
with respect to the previous cases. Thus, we generated $10^3$ 
samples of $N=10^8$ events. The results of the simulations are shown
in Fig.~\ref{fig:case3}, showing once again perfect  
agreement with the expectations, even if both the genuine and the
artificial modulations are present at the same time scale. 

\subsection{Case 4: isotropy with spurious effects and a broken East-West symmetry}

\begin{figure}[t]
\begin{center}
\subfigure[]
{\includegraphics[width=8.5cm]{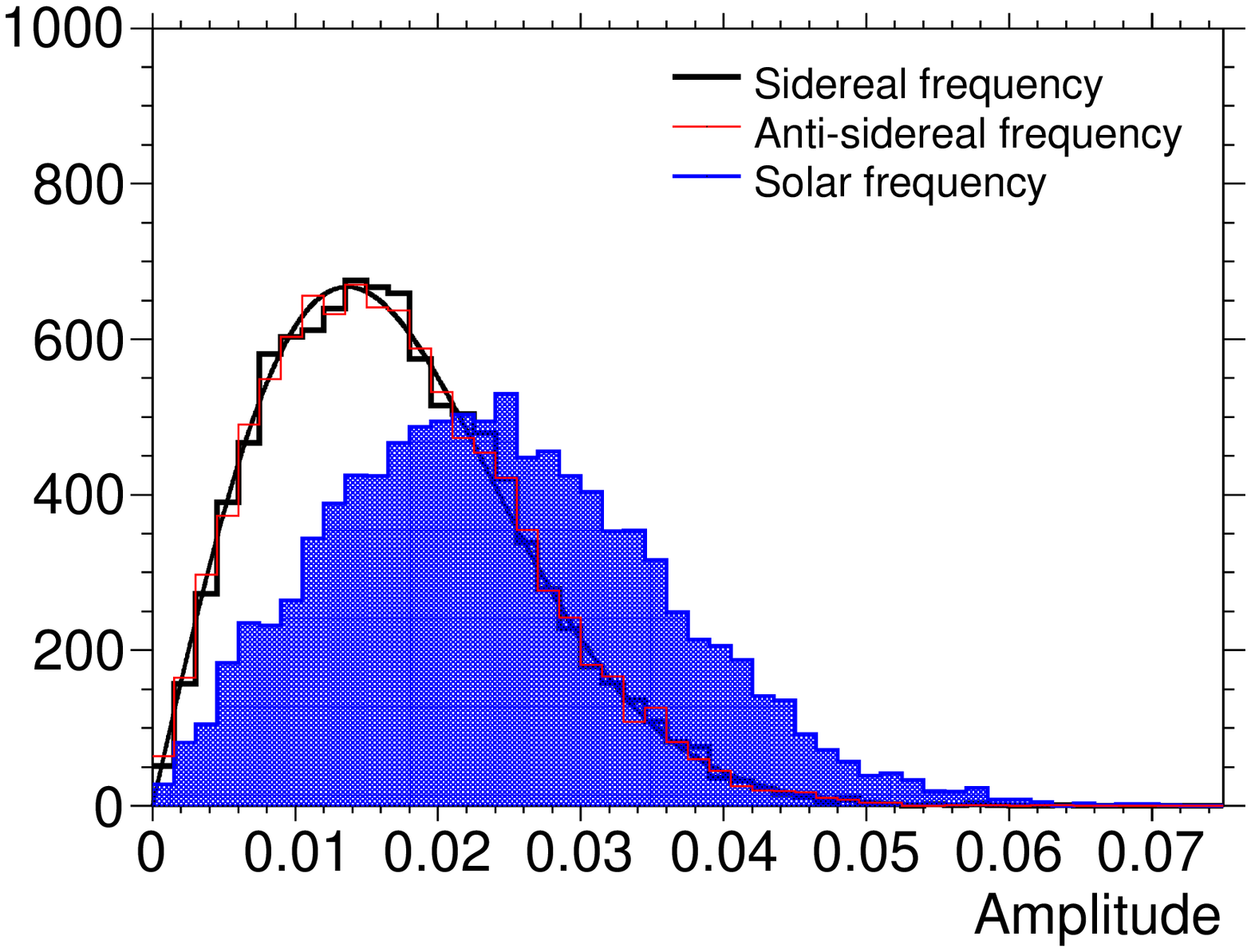}}
\subfigure[]
{\includegraphics[width=8.5cm]{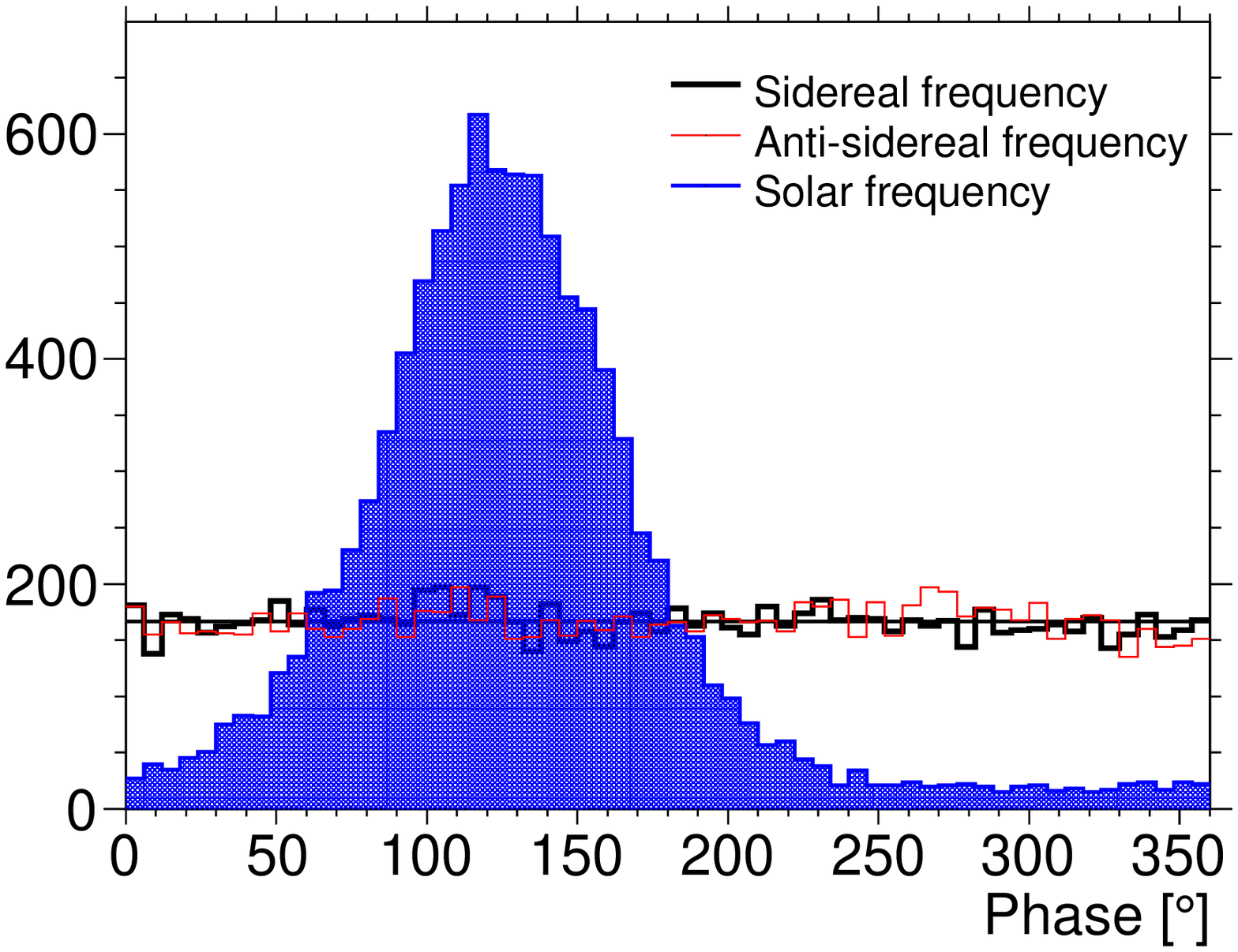}}
\end{center}\caption{Distribution of the amplitudes $(a)$ and phases $(b)$ of the first 
harmonic obtained with the East-West analysis in case of isotropy and of a broken East-West
symmetry of the detection efficiency of the experiment. Thick histogram: sidereal frequency.
Thin histogram: anti-sidereal frequency. Filled histogram: solar frequency.}  
\label{fig:case4}
\end{figure}

Finally, we study in this sub-section the influence of a broken East-West symmetry
of the detection efficiency of a ground array experiment. There are several reasons 
in practice why the East-West symmetry may be slightly broken. We consider here 
an array being slightly tilted, the tilt being specified by a unit vector $\hat{n}$
described by its zenith angle $\gamma$ (the \emph{tilt}) and its azimuth angle
$\phi_0$ (the \emph{direction}). If the unit vector $\hat{d}$ describes a given 
direction in the sky with zenith $\theta$ and azimuth $\phi$ the directional 
detection efficiency towards that direction is proportional to $\hat{d}\cdot\hat{n}$:
\begin{equation}
\hat{d}\cdot\hat{n}=\cos\gamma\cos\theta+\sin\gamma\sin\theta\cos(\phi-\phi_0)\simeq
\cos\theta\,\left[1+\gamma\tan\theta\cos(\phi-\phi_0)\right].
\end{equation}
Introducing this azimuthal dependence into the detection efficiency results in
a small difference between the Eastward and Westward counting rates, proportional to
$\gamma\cos{\phi_0}$ at first order in such a way that for any North/South asymmetry
the effect cancels exactly. Meanwhile, being \emph{independent of time}, this shift
does not impact itself in the estimate of the first harmonic. However, it is worth 
examining the effect of the \emph{combination} of the tilted array together with the 
modulations induced by weather effects on EAS developments, because this 
combination leads to an East-West counting rate proportional to $\gamma\eta(t)\cos{\phi_0}$, 
which may mimic a real East-West first harmonic modulation at the solar frequency. 

Generating $10^5$ isotropic samples of $N=10^4$ events on a tilted ground array
with a large value $\gamma=5^\circ$ and in the direction 
$\phi_0=0^\circ$ (leading to the maximal effect), we show in Fig.~\ref{fig:case4}
the results obtained at the sidereal frequency (thick histogram), the anti-sidereal
frequency (thin histogram), and the solar frequency (filled histogram). It is clear
that in such a case, only the solar frequency is affected by the East-West 
asymmetry introduced by the tilted array. On the contrary, the analyses performed
at the sidereal frequency remains unbiased.

\section{Conclusions} \label{section:conclusion}

The differential East-West method for the measurement of large scale anisotropies has 
been revisited. Using the fact that the experimental instabilities 
simultaneously affect both the East and the West sectors, 
we have shown that with this method the equatorial 
component of the dipole can be recovered in an 
unbiased way, without applying any corrections to account for spurious 
effects, but with a reduced sensitivity with respect to the standard Rayleigh
analysis. Despite of this reduced sensitivity, this method has the advantage of 
avoiding the need to correct the total counting rate for instrumental and
atmospheric effects. Finally, we have also shown that this method leads 
to unbiased results at the sidereal frequency even in the case of a broken 
East-West symmetry.

\vspace{1cm}
This article is devoted to the memory of \emph{Gianni Navarra}. Besides being a 
great scientist, he was deeply involved in this analysis and was the real inspirer 
of this work.
\vspace{0.4cm}

\section*{Acknowledgments}
We thank members of the Pierre Auger collaboration for useful
discussions, in particular Paul Sommers.
\textit{V.V.A.} is grateful to the INFN Gran Sasso National Laboratory
for financial support through FAI funds. \textit{P.L.G.} acknowledges the 
financial support by the European Community 7th
Framework Program through the Marie Curie Grant PIEF-GA-2008-220240. The work of \textit{S.M.} 
and \textit{E.R.} is partially supported by grants ANPCyT PICT 1334 and CONICET PIP 01830.

\end{document}